\documentclass[aps,prl,showpacs]{revtex4}
\usepackage{psfig}
\begin{document}        %
\draft
\title{Note on "Anomalous Hydrodynamic Drafting of \\ Interacting Flapping Flags"}
\author{Chu Z. K. Hua} 
\affiliation{9, Road Qingshan, Qitaizhonchang, Qitai 831809,
China}
\begin{abstract}
We make remarks on  Ristroph and Zhang's [{\it Phys. Rev. Lett.}
101, 194502  (2008)] paper.
%
\end{abstract}
\maketitle
\bibliographystyle{plain}
Ristroph and Zhang just showed that [1], inverted drafting could
be produced by flexible flags (of which flapping in front reduces
fluid forces). As reported in their experiments on {\it schooling}
flapping flags, they found that it is the leader of a group who
enjoys a significant drag reduction (of up to 50\%), while the
downstream flag suffers a drag increase. Thus, they remarked that
this counterintuitive inverted drag relationship is rationalized
by dissecting the mutual influence of shape and flow in
determining drag [1]. \newline The present author, however, likes
to issue some comments here about Ristroph and Zhang's
experimental procedure and data. Firstly, the technique for
measurements of the drag on a flag originated from [2] (cf. Ref.
15: {\it The flagpole is fixed to a cantilever which bows slightly
($<0.5$mm) under the fluid forcing of the flag. The deflection is
measured optically} in [1]; relating support deflection to drag).
However, is the net deflection only due to the net force?  We know
that once there is a net moment (torque included) applied to one
end of a (flat) cantilever there will be also a net deflection
[3]. Meanwhile, can the net {\bf thrust} (generated) [4] be
distinguished from the net drag in [1-2]? \newline Note that in
verifying the drag measuring technique [2] the flow is presumed to
be described by the inviscid, incompressible Euler equations [2]
by neglecting possible effects of flow compressibility due to
thickness variations in the soap film (skin friction or {\it
viscous drag} was also neglected [2]). The latter (compressibility
[5] as well as variations of film's thickness [6]) is still being
argued [7] and there is a possibility that {\bf Marangoni} flows
(due to surface-tension-driven mechanical instability) occur [8]
and will induce errors to the experimental measurements reported
in [1]. According to [5], the sound speed in the film  is $\sqrt{2
E_m/\rho_f d_f}$ $\approx 4.3$ m/s ($E_m$ : Marangoni elasticity
modulus ($=0.088$ N/m [5]), $\rho_f$ : fluid density, $d_f(=4.7
\mu$m [1]) : film thickness). The flow velocity ($U$) is $2$ m/s
in [1] and then the Mach number is around 0.46 which means the
flow is compressible.  The theoretical validation for those
relating bending deflection to drag in [1-2] which was based on
the incompressible flow thus should be checked again and possibly
be modified.
\newline Meanwhile as noted in Ref. 15 of [1], the {\it minute} bending of the
support  is optically measured by the same procedure in  [2]. How
can {\it The flat cantilever suppresses lateral motion, and
streamwise force fluctuations are damped by a viscous dashpot
attached to the beam} (cf. Ref. 15 in [1]) be still  valid during
large-amplitude flapping motions?
We know that the identification of an (equilibrium) neutral axis
(n.a.) [3] is crucial to the judgement of the bending behavior of
a beam (positive or negative deflection?). Can this n.a. be easily
found for largely fluctuating support? \newline The other argue is
about the fixed width (9.5 cm) of the planar water tunnel in [1]?
How about the {\bf interference} or induced {\bf blockage} [9]
(e.g. 4th. flag in Fig. 4a of [1]) between bodies, wakes, and
edges of films when large-amplitude flapping occurs (cf. Fig. 1)?
What happens once the width (soapy water descends in-between)
increases or decreases a little? Is 9.5 cm optimal for the
conclusion made in [1] (especially for smaller separations :
$G/L$, cf. [1] for $G$,$L$ details, where reported significant
anomalous inverted drafting appears; cf. Figs. 2 and 4 in [1]).
Finally, considering the limitation for the optical resolution in
[1] : can the total drag and drag increment be measured separately
for $G/L=0$? How to calibrate the elastic response of the support
instantaneously? These issues will dominate the conclusion made in
[1].
{\it Acknowledgements.} The author thanks Ms. Chu Mary and Hsieh
Jiu Xiang for their support.

\psfig{file=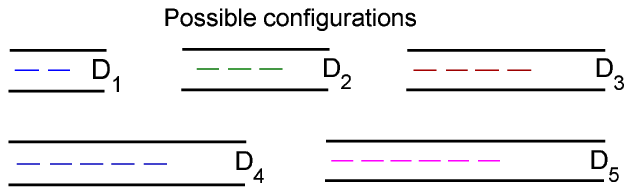,bbllx=-2cm,bblly=22.8cm,bburx=8cm,bbury=25cm,rheight=2.2cm,rwidth=6cm,clip=}
%
\begin{figure}[h]
\hspace*{1mm} Fig. 1 \hspace*{1mm} Schematic set-up for different
drag-force baseline calibration of tandem flags.  $D_0$ is for an
isolated flag [1]. Ristroph and Zhang neglected the differences in
[1] by presuming a universal $D_0$ for the normalization and
baseline comparison even the corresponding configurations are
different (say, two and six tandem flapping flags).
\end{figure}

\end{document}